%% file: jpegai_main.tex
\documentclass[a4paper,conference]{IEEEtran}
\IEEEoverridecommandlockouts

\usepackage{cite}
\usepackage{ntheorem}
\usepackage{algorithmic}
\usepackage{graphicx}
\usepackage{textcomp}
\usepackage{xcolor}
\usepackage{hyperref}
\usepackage{mathtools}
\usepackage{dirtytalk}
\usepackage{multirow}
\usepackage{comment}

\usepackage{url}
\usepackage{mwe} 
\usepackage{float}
\usepackage{comment}
\usepackage{tikz}
\usepackage{fancyvrb}
\usepackage{booktabs}
\usepackage{siunitx}
 \usepackage{epsfig}
 \usepackage{amsmath}
\usepackage{amssymb}

\usepackage{hyperref}
\usepackage{booktabs}

\usepackage[subpreambles=true]{standalone}
\usepackage[version=4]{mhchem}
\usepackage{lipsum}
\usepackage{tabularx}
\usepackage{tikz}
\usepackage{pgfplots}
\pgfplotsset{compat=1.18}
\usetikzlibrary{positioning}
\usepackage{float}
\usepackage[acronym]{glossaries}
 \glsdisablehyper
\loadglsentries{acronyms}

\def\BibTeX{{\rm B\kern-.05em{\sc i\kern-.025em b}\kern-.08em
    T\kern-.1667em\lower.7ex\hbox{E}\kern-.125emX}}

\begin{document}
\title{Subjective Visual Quality Assessment for High-Fidelity Learning-Based Image Compression \\[5mm]

\thanks{This research is funded by the DFG (German Research Foundation) -- Project ID 496858717, titled “{JND}-based Perceptual Video Quality Analysis and Modeling”. D.S. is funded by DFG Project ID 251654672.}
}
\author{\IEEEauthorblockN{
Mohsen Jenadeleh\IEEEauthorrefmark{1},
Jon Sneyers\IEEEauthorrefmark{2},
Panqi Jia \IEEEauthorrefmark{3},
Shima Mohammadi\IEEEauthorrefmark{4},
João Ascenso\IEEEauthorrefmark{4},
 Dietmar Saupe\IEEEauthorrefmark{1}}
\IEEEauthorrefmark{1}University of Konstanz, Germany \hspace{0.7em}
\IEEEauthorrefmark{2}Cloudinary, Belgium \hspace{0.7em}
\IEEEauthorrefmark{3}Huawei, Germany\hspace{0.7em}
\IEEEauthorrefmark{4}IST-IT, Portugal
\\

\{mohsen.jenadeleh, dietmar.saupe\}@uni-konstanz.de, jon@cloudinary.com, \\ 
panqi.jia@huawei.com , \{shima.mohammadi, joao.ascenso\}@lx.it.pt}
\IEEEoverridecommandlockouts
\IEEEpubid{\makebox[\columnwidth]{
\copyright 2025 IEEE \hfill} \hspace{\columnsep}\makebox[\columnwidth]{ }}

\maketitle


\begin{abstract}
Learning-based image compression methods have recently emerged as promising alternatives to traditional codecs, offering improved rate-distortion performance and perceptual quality. JPEG AI represents the latest standardized framework in this domain, leveraging deep neural networks for high-fidelity image reconstruction. In this study, we present a comprehensive subjective visual quality assessment of JPEG AI-compressed images using the JPEG AIC-3 methodology, which quantifies perceptual differences in terms of Just Noticeable Difference (JND) units.
We generated a dataset of 50 compressed images with fine-grained distortion levels from five diverse sources. A large-scale crowdsourced experiment collected 96,200 triplet responses from 459 participants. We reconstructed JND-based quality scales using a unified model based on boosted and plain triplet comparisons. Additionally, we evaluated the alignment of multiple objective image quality metrics with human perception in the high-fidelity range. The CVVDP metric achieved the overall highest performance; however, most metrics including CVVDP were overly optimistic in predicting the quality of JPEG AI-compressed images.
These findings emphasize the necessity for rigorous subjective evaluations in the development and benchmarking of modern image codecs, particularly in the high-fidelity range. Another technical contribution is the introduction of the well-known Meng–Rosenthal–Rubin statistical test to the field of Quality of Experience research. This test can reliably assess the significance of difference in performance of quality metrics in terms of correlation between metrics and ground truth. The complete dataset, including all subjective scores, is publicly available at \href{https://github.com/jpeg-aic/dataset-JPEG-AI-SDR25}{https://github.com/jpeg-aic/dataset-JPEG-AI-SDR25}.
\end{abstract} 
\begin{tikzpicture}[overlay, remember picture]
\path (current page.north) node (anchor) {};
\node [below=of anchor] {%
2025 17th International Conference on Quality of Multimedia Experience (QoMEX)};
\end{tikzpicture}
\vspace{-20pt}
\section{Introduction}
\label{sec:introduction}
Image compression remains a fundamental research area in image processing, having undergone significant advancements over the years. Traditional image compression standards—such as JPEG~\cite{jpeg}, JPEG~2000~\cite{rabbani2002overview},  AVIF~\cite{barman2020evaluation}, HEIC~\cite{sullivan2012overview}, VVC/H.266 intra coding \cite{hamidouche2022versatile}, and JPEG~XL~\cite{alakuijala2019jpeg} are designed to reduce data redundancy while preserving visual fidelity. These codecs use hand-engineered transformations, including the discrete cosine transform (DCT) and discrete wavelet transform (DWT), followed by quantization and entropy coding. More recently, deep learning-based image compression has emerged as a promising alternative, leveraging neural networks to further optimize compression efficiency with a perceptual quality target \cite{hu2021learning,cheng2018deep}. 
Unlike traditional codecs, learning-based methods employ end-to-end trainable architectures for the encoding and decoding processes and have demonstrated enhanced adaptability, enabling the preservation of critical visual details while achieving lower bitrates. Furthermore, these approaches facilitate adaptive quantization, content-aware bitrate allocation, and more effective entropy modeling, positioning them as viable solutions for the evolving challenges in modern image compression.

One of the most recent advancements in the field is the development of the JPEG~AI standard \cite{ascenso2023jpeg,alshina2024jpeg,jia2024bit}, a state-of-the-art image compression standard being developed by the Joint Photographic Experts Group (JPEG). Unlike conventional transform-based codecs such as JPEG, JPEG 2000, and JPEG XL, this new standard employs deep learning-based image coding techniques to learn optimal encoding and decoding strategies. By leveraging neural network-driven models, JPEG~AI achieves higher compression efficiency while maintaining superior visual fidelity, signaling a transformative shift towards AI-powered end-to-end image compression. 
However, this type of methods generate novel types of artifacts, distinct from traditional blocking and ringing distortions, while achieving competitive performance compared to conventional approaches \cite{ascenso2020learning}. Objective metrics such as SSIM \cite{ssim}, MS-SSIM \cite{ms-ssim}, PSNR, and VMAF \cite{vmaf} 
offer some insights and have been used by researchers to evaluate compression algorithms \cite{yu2022evaluating,mentzer2018conditional,cheng2020learned,testolina2021performance}. However, the artifacts produced by learning-based codecs necessitate comprehensive subjective studies to assess their impact on perceived image quality \cite{mohammadi2023fidelity,cheng2019perceptual}.
\begin{figure*}[ht!]
\centering
    \includegraphics[width=0.9\linewidth,height=2.5cm]{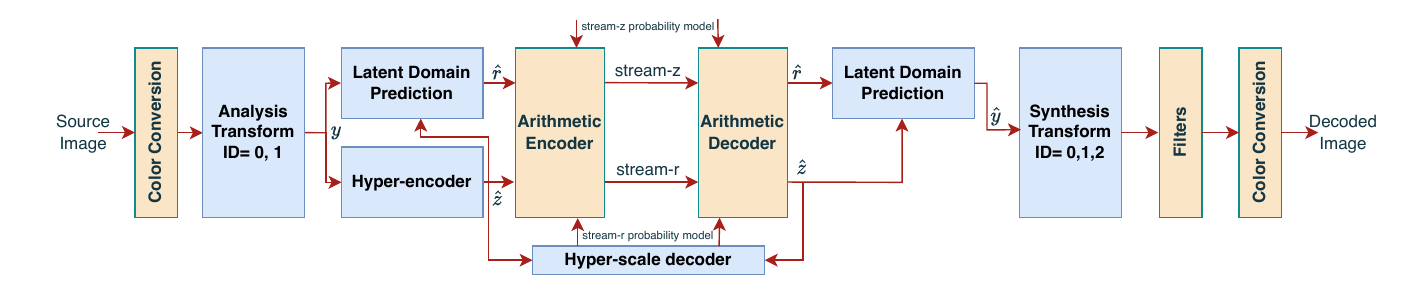}
        \vspace{-10pt}
    \caption{JPEG~AI encoder and decoder architecture (blue modules correspond to neural networks).}
    \label{fig_jpeg_ai}
    \vspace{-5pt}
\end{figure*}
Despite the increasing prevalence of learning-based image compression, relatively few studies have focused on its subjective quality assessment. In \cite{cheng2019perceptual}, a subjective study was conducted using the absolute category rating (ACR) method, in which seven source images were compressed with image learning compression solution at four different bitrates—ranging from very low to very high—using three early learning-based compression algorithms. Similarly, in \cite{mohammadi2023fidelity}, a subjective evaluation methodology based on JPEG~AIC-2 Annex A \cite{iso29170-2} and a triplet comparison approach was employed to assess subjects' preferences between images compressed with two learning-based algorithms, LBIC-CO and LBIC-PO. This study included 46 source images, each encoded at five different bitrates, and collected ratings from 20 subjects. Additionally, in \cite{li2021quality}, a dataset comprising 100 source images with varying resolutions was constructed, where images were compressed using three traditional codecs and seven learning-based compression algorithms, each at three or four bitrates, ranging from very low to very high quality. A double-stimulus method with a five-category degradation rating scale was then used to collect responses from 40 subjects.

In all of these studies, subjective quality assessment relied on single- or double-stimulus category ratings or user preference. However, as interest grows in the perceptual evaluation of high-fidelity compressed images, new methodologies are being introduced. To address this, JPEG~AIC-3 has proposed a subjective test methodology for estimating the perceptual quality of compressed images, particularly in the high-quality to perceptually lossless range, using Just Noticeable Difference (JND) units. In \cite{testolina2024fine}, the JPEG~AIC-3 methodology was applied to evaluate the performance of traditional codecs, including JPEG, JPEG 2000, AVIF, VVC, and JPEG-XL. In this study, the JPEG~AIC-3 subjective test methodology was applied to conduct a large-scale crowdsourcing study of JPEG~AI compressed images with fine-grained distortion levels and reconstructed the subjective scores in JND units.

The main contributions of this work are:
\begin{itemize}
    \item A dataset of JPEG~AI-compressed images and triplet comparisons with plain and boosted distortions according to JPEG AIC-3.
    \item A large-scale crowdsourcing study with 459 participants.
    \item Data analysis of JPEG AIC-3 by subject screening, outlier detection and handling, and maximum likelihood estimation of an exponential unified model for perceived plain and boosted distortion in JND units.
    \item Performance evaluation of 15 full-reference image quality assessment (IQA) metrics on our dataset.
    \item Introduction of the Meng-Rosenthal-Rubin statistical test to assess the significance of differences in the correlations of quality metrics with ground truth.
\end{itemize}

\section{JPEG~AI standard}

The scope of JPEG~AI standardization \cite{ascenso2023jpeg} is the creation of a learning-based image coding standard offering a single-stream, compact compressed domain representation, targeting both human visualization, and effective performance for image processing and computer vision tasks, with the goal of supporting a royalty-free baseline. The standardization process is divided into two versions, where version 1 focuses on high perceptual quality and fidelity, reconstructing images through entropy decoding and image synthesis from a latent tensor representation. This has been the main target until now. The International Standard (IS) for JPEG~AI Part 1 (Core Coding Engine) is on publication phase \cite{ISOJPEGAI-1} and will be made available soon. Work is also underway on JPEG~AI profiles and levels (Part 2), reference software (Part 3), conformance testing (Part 4), and file format specifications (Part 5). JPEG~AI employs a multi-branch decoding framework \cite{alshina2024jpeg}, allowing a single codestream to be reconstructed in multiple ways, each with different trade-offs between complexity and quality. This adaptability ensures broad support across several devices and applications. After entropy decoding retrieves quantized residual samples and reconstructs latent samples, the core decoding engine defines three synthesis (inverse) transforms, each capable of producing a reconstructed image. Additionally, conformance testing, still in development, explores the possibility of standard-compliant decoding without requiring bit-exact reconstruction. By supporting multiple synthesis transforms and providing flexibility in reconstruction accuracy, JPEG~AI enables vendors to optimize implementations to best suit their device capabilities and application needs.

The high-level diagrams of the JPEG~AI encoder and decoder are shown in Fig. \ref{fig_jpeg_ai}. As usual, the JPEG~AI standard defines encoder operations as non-normative, included only to facilitate understanding of the normative decoder operations, which includes weights and other parameters. The encoder starts by converting the source image to the YUV color space as defined in the BT.709 standard, the format internally supported by the JPEG~AI codec. This involves separating the image into primary and secondary color components, both of which undergo the same compression steps: analysis transform, latent domain prediction, hyper-encoding, and residual coding using an arithmetic encoder (AE). The analysis transform uses convolutional and non-linear activation layers to decorrelate the source image, producing a latent representation, $y$. Two possible synthesis transforms are described in the standard, with and without attention model. This latent representation is further processed into a very compact hyper-tensor, $z$, which is encoded before residual computation to enable efficient latent domain prediction and the creation of the entropy coding probability model. The hyper-tensor $z$ is quantized to $\hat{z}$ and compressed using an arithmetic encoder with probability model obtained from the trained model, which is shared between the encoder and decoder. Latent domain prediction then computes a residual $r$, which subtracted from a prediction obtained from $y$ and then quantized (rounded). This residual is encoded using an arithmetic coder with entropy parameters derived from the hyper-scale decoder, producing stream-r (see Fig. \ref{fig_jpeg_ai}). 

The decoder operations mirror those of the encoder in reverse order. First, stream-z is parsed, and the hyper-scale decoder generates the entropy probabilistic model, which provides the parameters for residual decoding. Note that this operation is performed at encoder and decoder to have exactly the same model at both sides. Next, stream-r is parsed, and residual $r$ is recovered through arithmetic decoding. Following this, latent domain prediction is performed using $\hat{z}$ with a hyper-decoder and a multistage context model, leveraging previously decoded information. Finally, one of the three synthesis transform aforementioned outputs the decoded image. The primary component is processed independently, while for the secondary component, it incorporates latent representations from  the primary and secondary components as auxiliary input.

\section{JPEG~AI compressed image dataset}
\newcommand{\putpng}{\includegraphics[height=4.0cm]}
\newcommand{\putpngt}{\includegraphics[height=2.2cm]}
\newcommand{\commentout}[1]{}

 \begin{figure*}[ht!]
    \centering
    \setlength{\tabcolsep}{3pt}
    \begin{tabular}{ccccc}
    \putpngt{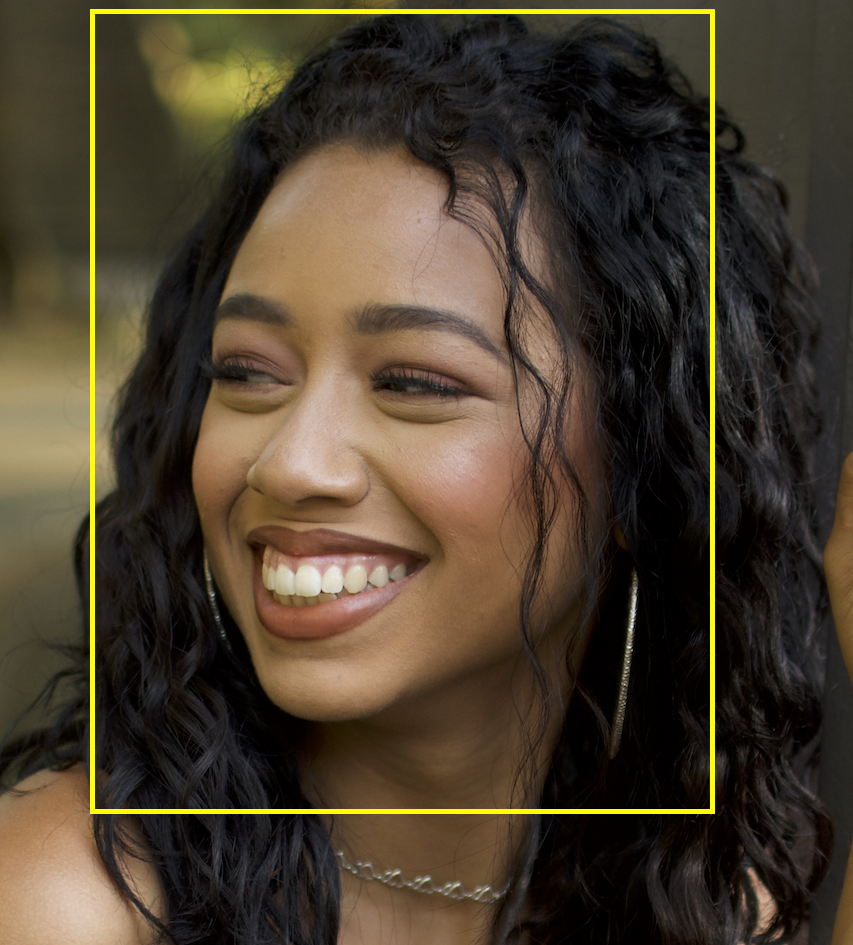}
    & \putpngt{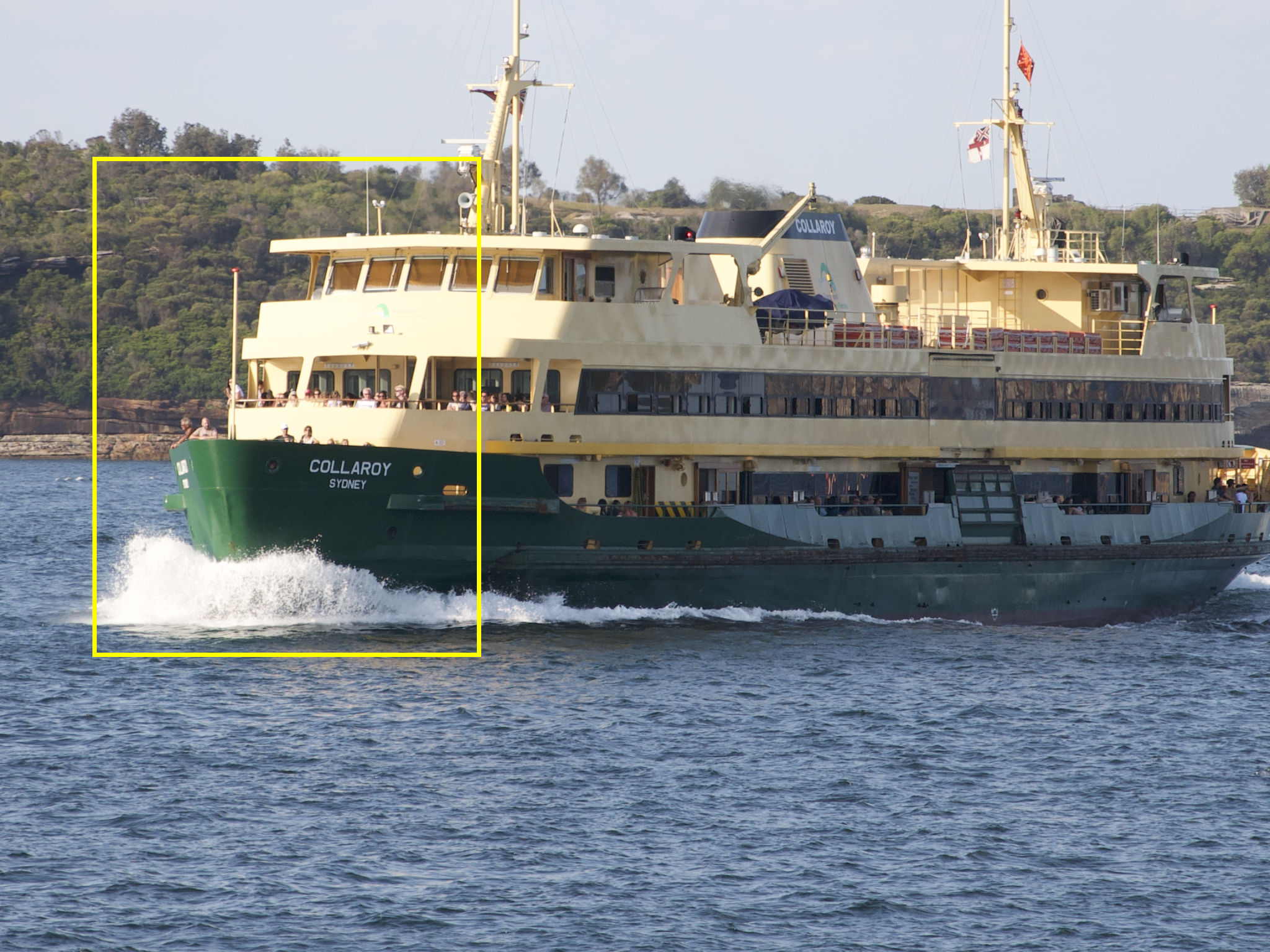}
    & \putpngt{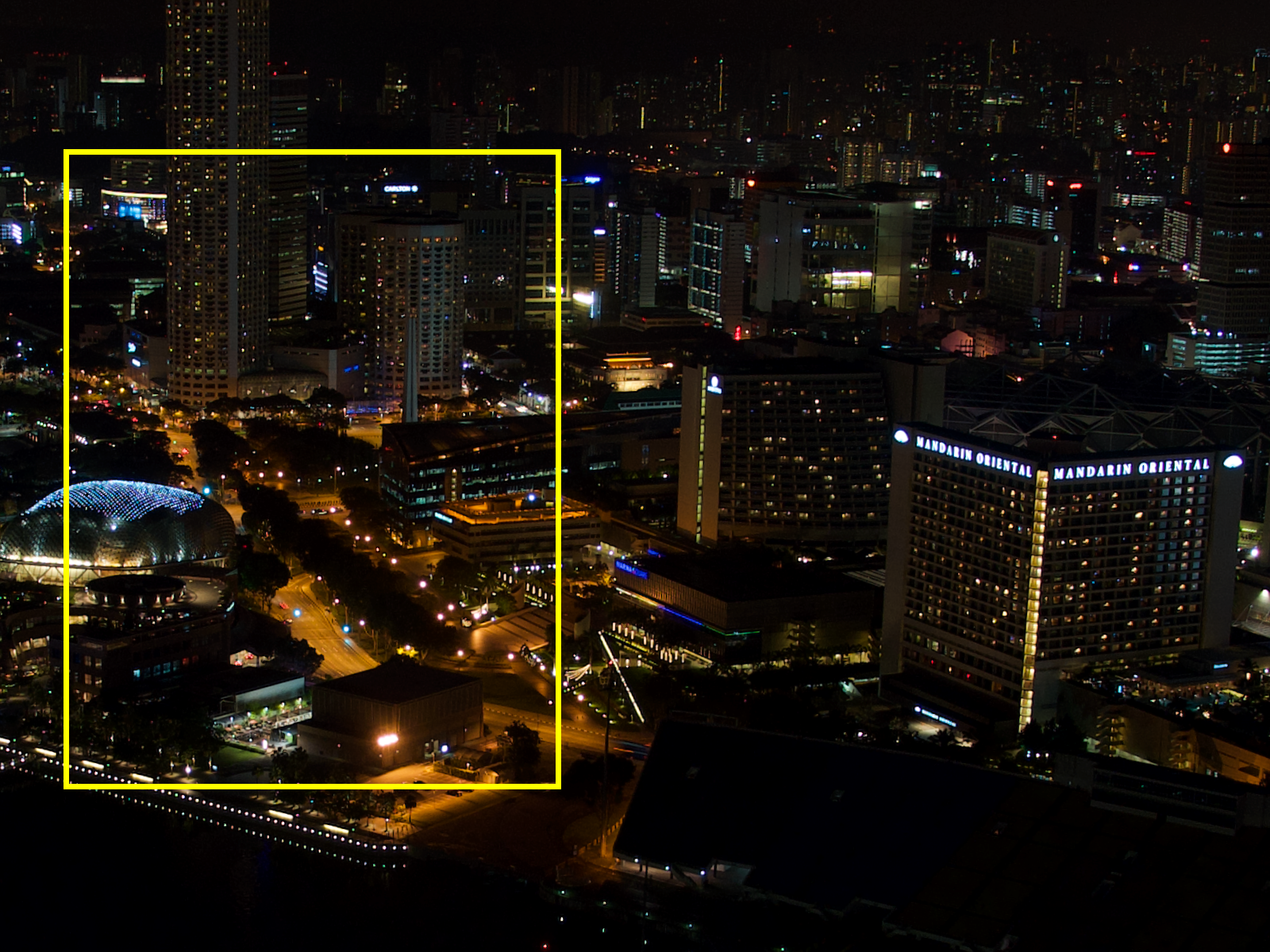}
    & \putpngt{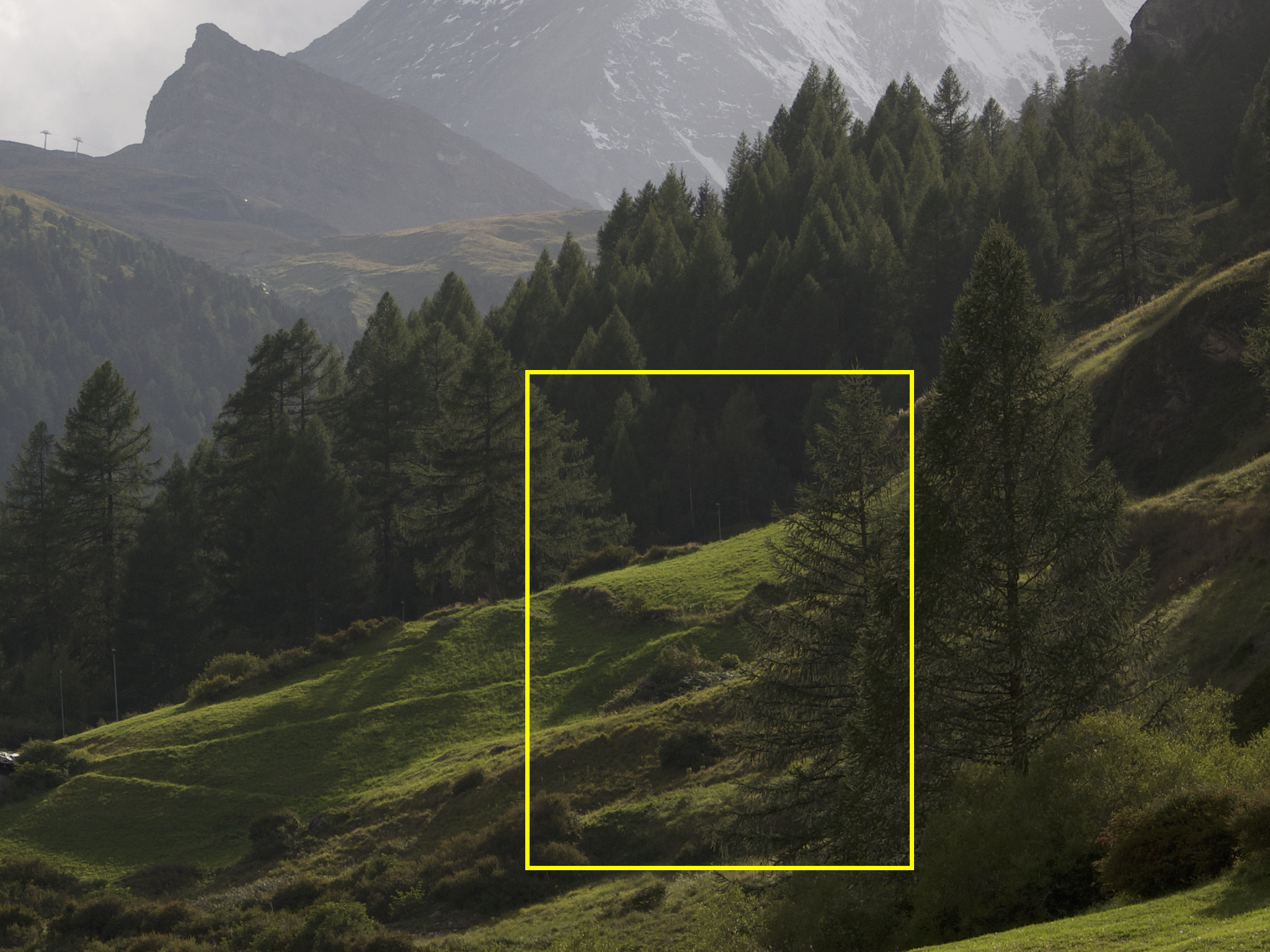}
    & \putpngt{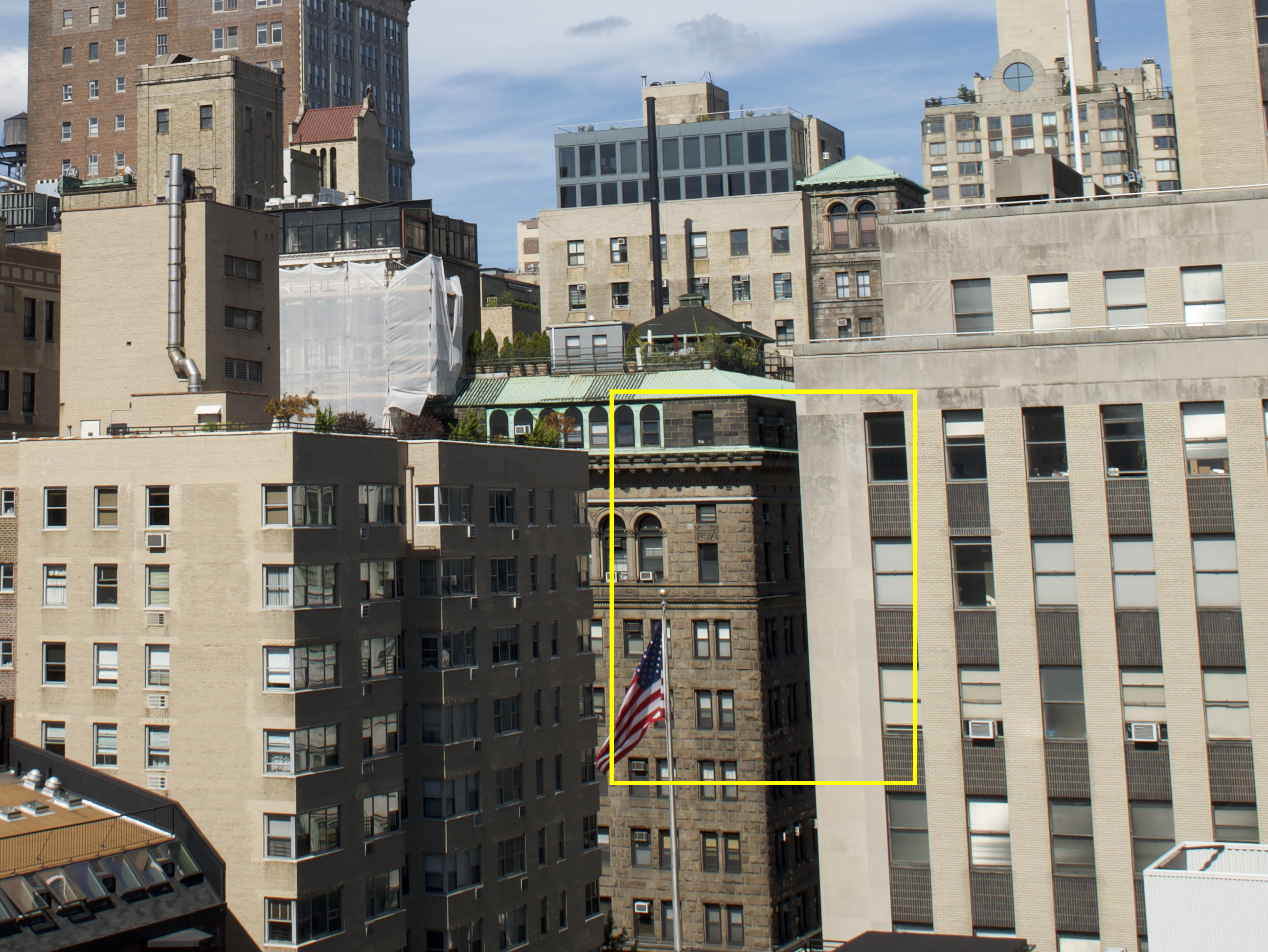}
    \\
    \putpng{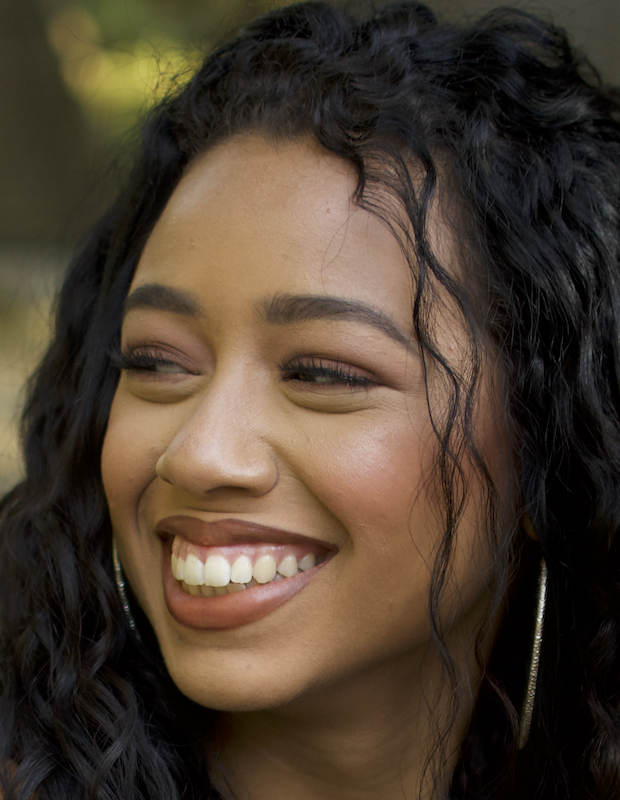}
    & \putpng{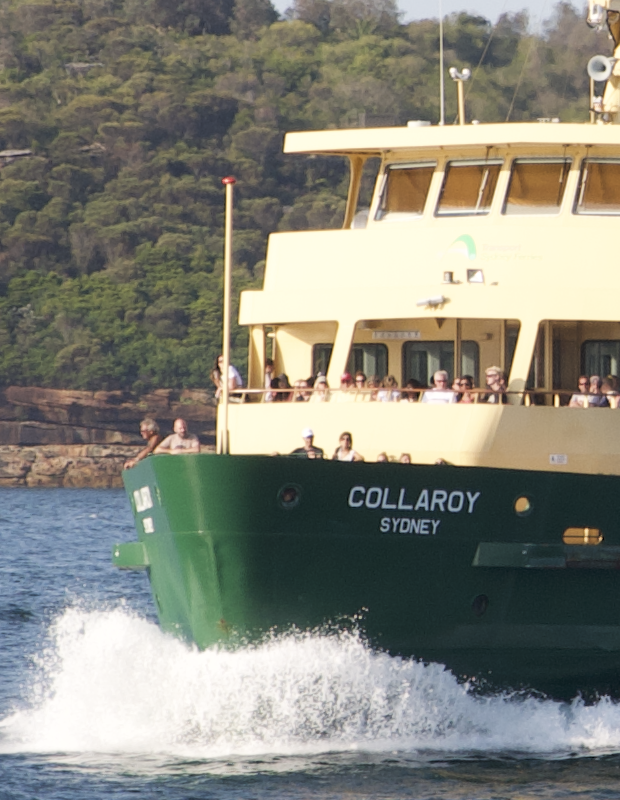}
    & \putpng{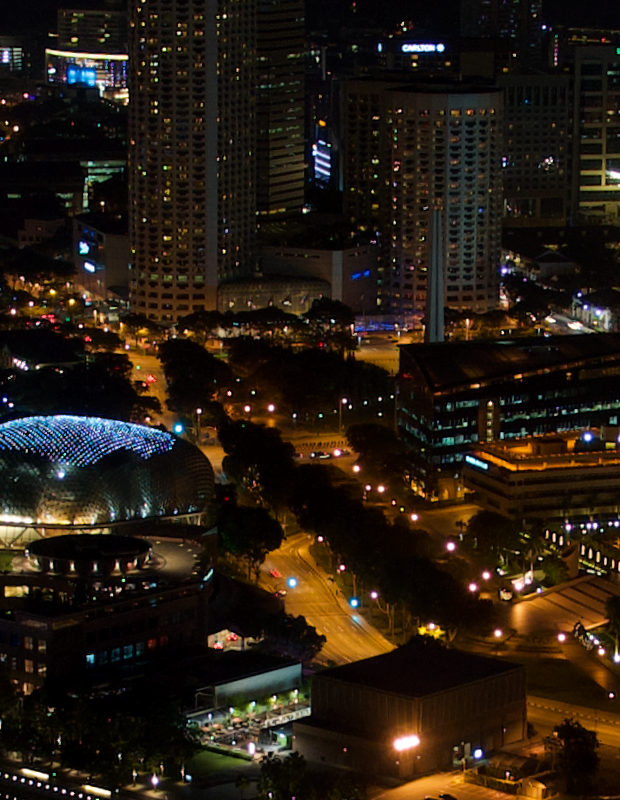}
    & \putpng{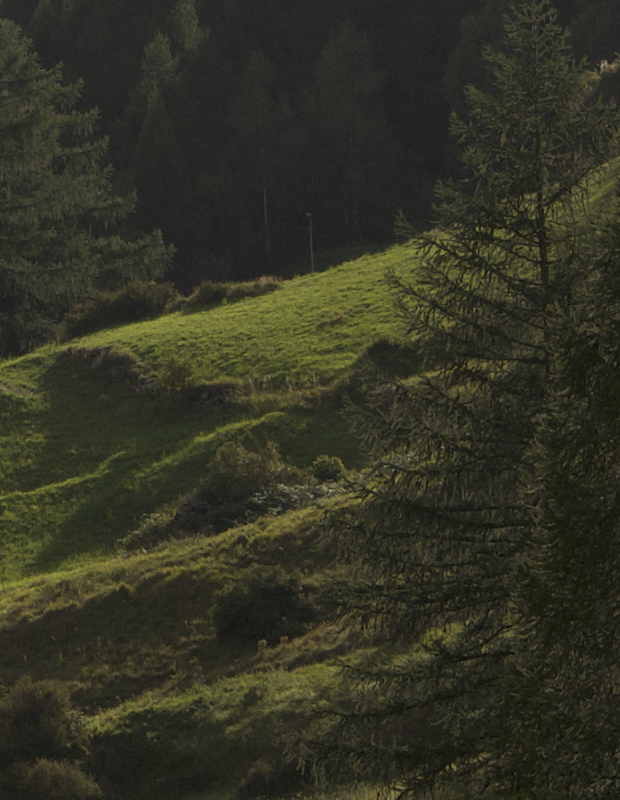}
    & \putpng{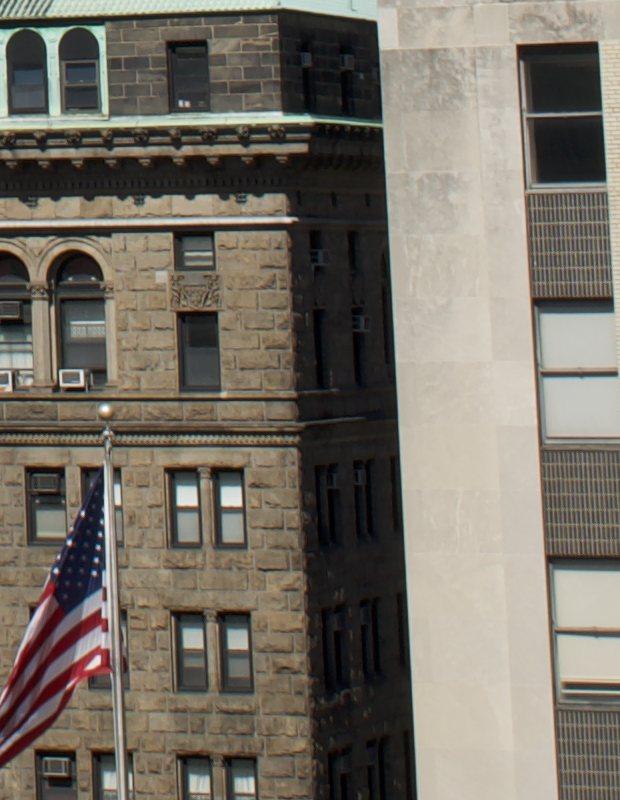}
    \\
    00002 & 00006 & 00007 & 00009 & 00010
    \end{tabular}%
    \caption{Full-resolution source images (top) and their cropped versions (bottom).  Crops were extracted at 620$\times$800 resolution, with upper-left coordinates listed in Table~\ref{tab_dataset}, which also provides the content category and full-resolution dimensions of each source. 
    The cropping was designed to preserve key visual features for crowdsourcing assessments.}
        \label{fig_dataset}
\end{figure*}
 \subsection{Source images}
The five source images used by the JPEG~AIC group in their recent work \cite{testolina2024fine} to evaluate their subjective test methodology for fine-grained image quality assessment were used. These images are shown in Fig. \ref{fig_dataset}. The selected source images, taken from the JPEG~AIC-3 dataset \cite{testolina2023jpeg}, were chosen to represent diverse image types and content at different resolutions. For crowdsourced image quality assessment, the JPEG~AIC group manually selected an interesting region from each source image and cropped it to $620\times800$ pixels. The cropped regions were chosen to retain key structural details and visual complexity, making them representative of the distortions that would be perceived in the full-resolution images. Table \ref{tab_dataset} summarizes these five source images.  
 
\begin{table}[t]
\caption{Source images and cropping details (see Fig.~\ref{fig_dataset}).}
\vspace{-8pt}
\label{tab_dataset}
\centering
\begin{tabular}{c c c c}
\toprule
Source image & Content & Resolution & Crop at (x,y)  \\ \midrule[0.1em]
 00002 & Human face & 853$\times$945 & (92,11)   \\
 00006 & Scene with water & 2048$\times$1536 & (152,256)  \\
00007 & Night scene & 1600$\times$1200 & (83,191)  \\
 00009 & Landscape & 2048$\times$1536 & (850,600)  \\
00010 & Buildings & 2592$\times$1946 & (1250,800) \\  \midrule[0.1em]
\end{tabular}
\vspace{-20pt}
\end{table}

\subsection{JPEG~AI coding}

The JPEG~AI coding engine was set to the high operating point with all tools enabled, utilizing YUV444 as the internal color space. This configuration employs advanced analysis/synthesis transforms (IDs 0/2) with attention models. All switchable coding tools, including post-processing filters, were activated. These content-adaptive tools dynamically scale intermediate data (e.g., residuals) to enhance perceptual quality. To generate a range of decoded images, 10 rate points between 0.3 and 1.65 bpp were defined, using the three highest-rate models (out of four) in the JPEG~AI VM. The JPEG~AI VM7.0 was used with the command line: 

\begingroup
\scriptsize
\begin{verbatim}
python -m src.reco.scripts.eval --cfg ./cfg/tools_on.json 
./cfg/oper_point/hop.json ./cfg/BRM/regen_list.json 
--coding_type enc_dec -target_bpps [BPP*100]
\end{verbatim}
\endgroup

\subsection{Target bitrates selection }
In the JPEG~AIC study \cite{testolina2024fine}, five source images were compressed using five traditional codecs, namely, JPEG, JPEG 2000, VVC Intra, JPEG XL, and AVIF, at ten different bitrates, corresponding approximately to JND values evenly spaced between 0.25 and 2.5. For JPEG~AI, each source image was also compressed at ten distortion levels,
using bitrates between 0.3 bpp and 1.65 bpp increasing in steps of 0.15 bpp. It was visually checked that this range of bitrates roughly matches the perceived distortion ranges of the other codecs.

The JPEG~AIC-3 test methodology \cite{testolina2024fine} uses boosting techniques namely boosted triplet comparisons (BTC) to enhance the visibility of subtle distortions. These include zooming, where the plain images are cropped to half their size and upscaled using Lanczos resampling; artifact amplification, which scales the pixel-wise difference between the original and distorted images by a factor of 2 in each color channel; and flicker effect, where the reference and distorted images alternate at 10 Hz, each displayed for 100 ms per cycle. This methodology compares triplets in both plain triplet comparisons (PTC) and the BTC formats. We generated the boosted version of each plain compressed image by applying zooming and artifact amplification. The flickering technique is implemented in JavaScript and applied in real-time when the image triplets are shown to the participants.

\section{Experimental setup and procedure}
\subsection{Batch generation}
Triplets for BTC and PTC were generated following the procedure outlined in \cite{testolina2024fine}. Each triplet \((I_i, I_0, I_k)\) consists of two compressed images and the original source image.

\subsection{BTC and PTC Procedures}
In BTC, a test image alternates with its source at 10 Hz to induce a flicker effect. Observers identify the image with the most noticeable flicker or select ``Not sure'' if uncertain. In PTC, a toggle button allows observers to switch between the compressed and original images, with at least one toggle required before submitting a response. They were also limited to two toggle per seconds. BTC included all 10 distortion levels plus the source, while PTC was limited to five levels (2:2:10) plus the source. The comparisons comprised:

\begin{itemize}
    \item  Same-codec questions: Comparisons between images compressed with the same codec at different bitrates.
    \item  Cross-codec questions: Comparisons across codecs to align quality scales.
    \item  Trap questions: Pairs of the most distorted image (level 10) with its source  to detect unreliable subjects.
\end{itemize}

\subsection{Triplet distribution and study design}
Each source image yielded 110 BTC and 30 PTC same-codec triplets, with 20\% cross-codec triplets added. 

For the five source images, the BTC method included a total of 660 triplets, divided into five batches of 132 questions each. The PTC method consisted of 180 triplets, split into two batches of 90 questions each. To ensure response reliability, 10 trap questions were added to each batch.
\subsection{Crowdsourcing}
The same two web interfaces developed by JPEG~AIC-3 for BTC and PTC were used for this experiment. A screenshot of the interfaces are shown in the recent work of JPEG~AIC-3 \cite{testolina2024fine}. It was collected 49 responses per triplet for the PTC experiment and 120 responses per triplet for the BTC experiment. These numbers are selected to match the responses collected per triplet in \cite{testolina2024fine}. Participants were recruited through Amazon Mechanical Turk (MTurk) platform for the BTC and PTC experiments, which were conducted separately. Each participant could complete up to two different batches, with questions order is randomized for each participant. To ensure the required number of responses per triplet, 73 workers were recruited for the PTC experiment and 386 for the BTC experiment.  Experimental procedures were approved by the University of Konstanz  ethics committee.
 
\section{Experimental results}
\subsection{Data Cleansing}
\label{JPEG_AI_screens}
\begin{figure}[t!]
    \centering
    \scriptsize
    \setlength{\tabcolsep}{0pt}
    \begin{tabular}{cc}
         \includegraphics[trim={1.21cm 0 0cm 0},clip,width=0.48\columnwidth]{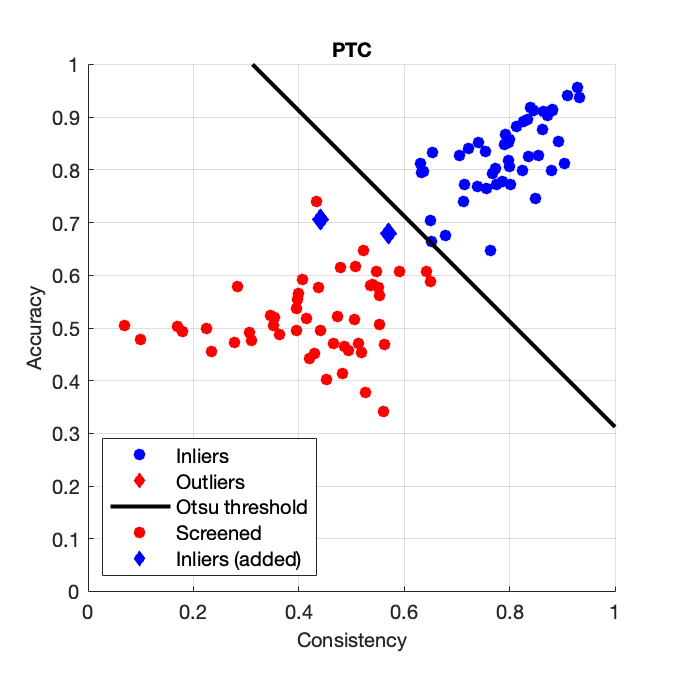}
   
  &       \includegraphics[trim={1.2cm 0 0cm 0},clip,width=0.48\columnwidth]{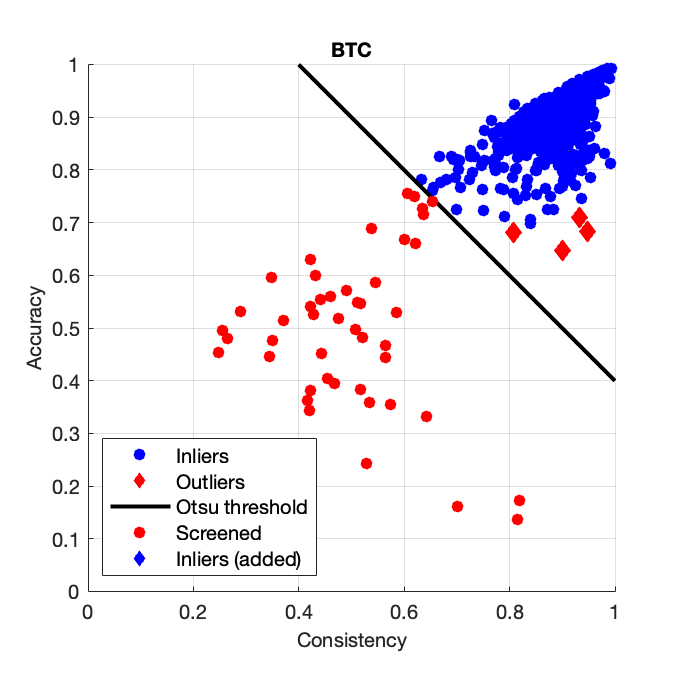}

    \end{tabular}
    \vspace{-10pt}
    \caption{Accuracy and consistency of batches for PTC and BTC.}
    \label{fig_filtering}
    \vspace{-13pt}
\end{figure} 
For reliability, batches of subjects were screened using the JPEG \mbox{AIC-3} method \cite{ISOJPEGAIC2024}, evaluating responses based on accuracy and consistency. Subjects scoring below threshold values were marked as screened.

\begin{figure*}[ht!]
\includegraphics[width=\linewidth]{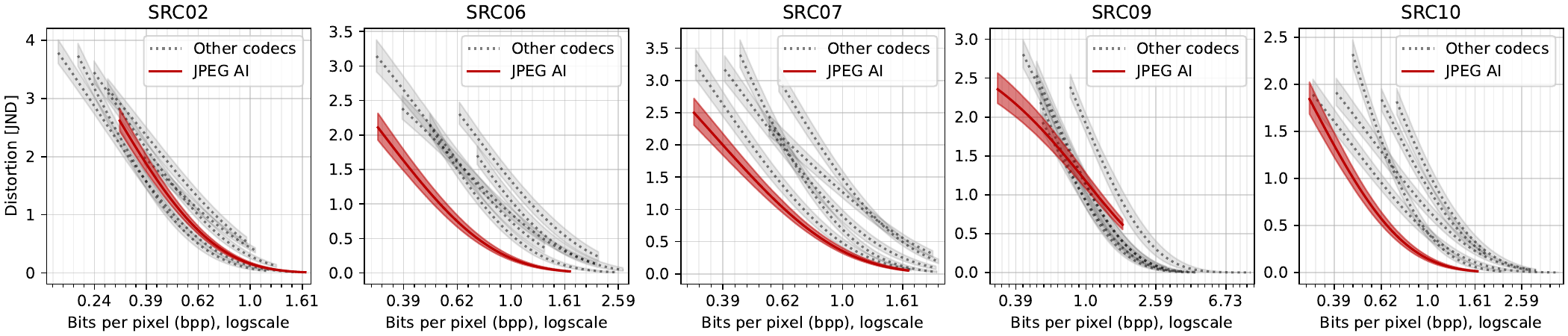}%
\vspace{-20pt}
\caption{Bitrate-distortion curves for the 5 source images. The shaded region indicates a 95\% confidence interval.}
\label{fig:bitrate_distortion}
\end{figure*}
Accuracy: This metric was computed using only comparisons where both images were encoded with the same codec. A response was deemed correct when the image with the lower bitrate was identified as more distorted. Responses labeled as ``Not sure'' contributed a score of 0.5. To reflect the perceptual strength of each trial, response contributions were weighted according to the magnitude of the distortion difference.

Consistency: Because the questions were presented in symmetric pairs, intra-batch consistency was evaluated by comparing responses across these matched pairs. A score of 1 was assigned when both responses were the same. If one response was ``Not sure'' while the other was not, the pair received a partial score of 0.375. Pairs with contradictory directional choices received a score of 0. The final consistency score was weighted based to the magnitude of the distortion difference.

Otsu thresholding of the average of accuracy and consistency of a batch was used. The thresholds were 0.6563 for PTC and 0.6992 for BTC, screening 51 of 98 PTC batch instances and 46 of 600 BTC batch instances of our dataset of JPEG~AI compressed image. 

Subsequently, an outlier detection method was applied according to \cite{ISOJPEGAIC2024} in which a few of the screened batches were exchanged with some of the others that gave a worse fit to the consensus of the inliers. In this process, 4 BTC batch instances were marked as outliers, and 2 screened PTC batch instances were relabeled as inliers. Fig.~\ref{fig_filtering} gives an overview of the screening and outlier detection for PTC and BTC.

\subsection{Model}

The reconstruction of perceptual quality scales followed the approach proposed by JPEG~AIC-3 \cite{ISOJPEGAIC2024}, briefly outlined here. The BTC and PTC responses were used together to reconstruct (boosted and non-boosted) scale values for the compressed images from all six codecs for all five source images. ``Not sure'' responses were split into half ``left'' and half ``right''. All responses were then interpreted in the sense of two-alternative forced choice in pair comparisons following the Thurstonian Case V model. Maximum likelihood estimation (MLE) yielded the coefficients of an exponential functional model ($d(r) = \alpha \exp{(-\beta r)}$) for the distortion-rate function of the non-boosted stimuli. Simultaneously, also the coefficients of a quadratic boosting transfer function ($t(d) = \gamma_1 d + \gamma_2 d^2$) were estimated that transforms the non-boosted scales $d(r)$ to the boosted ones, $t(d(r))$.\footnote{Note that this is different from \cite{testolina2024fine}, where we first independently estimated the pointwise scales for boosted and non-boosted stimuli and then aligned them using the quadratic boosting function and least-squares regression.}

The confidence intervals were obtained using $n = 1000$ bootstrap samples of the (filtered) BTC and PTC data by resampling with the replacement of the responses for each triplet question. The following scale reconstructions gave $n$ values for each bitrate in the corresponding ranges, which yielded the 95\% confidence intervals.

Fig.~\ref{fig:bitrate_distortion} shows the reconstructed bitrate-distortion (BD) curves for the JPEG~AI-compressed images. The curves for the other five codecs are shown in gray to illustrate general trends, although codec comparison is not the objective of this study. Also note that for source 09, due to the way crops were made, the subjects do not evaluate the same area of the image in BTC and PTC tests, which may result that the quality of the BTC crop cannot be extrapolated to the quality of the PTC crop (e.g. BTC quality lower than PTC). In any case, the confidence intervals (CIs) are narrow: for every codec including JPEG~AI, and for every source, the width of the CI of an image at $x$ JND is smaller than $0.1 + 0.05x$.

 \begin{figure*}[t!]
\centering

\centering
\begin{minipage}{0.19\linewidth}
  \includegraphics[width=3.5cm, height=3.0cm]{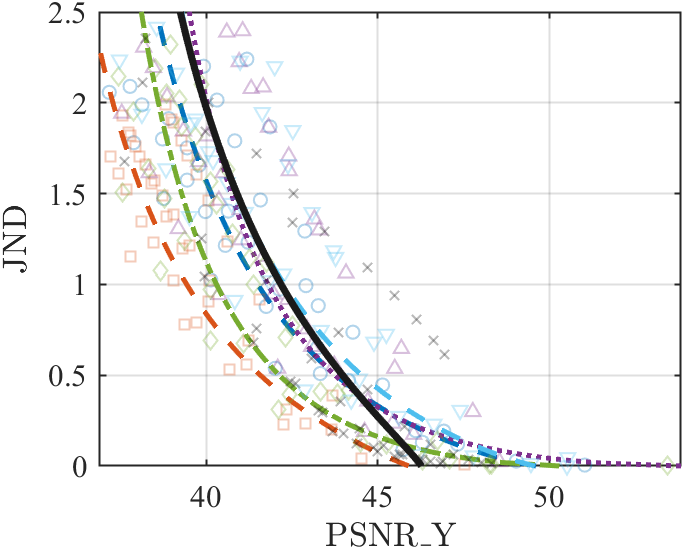}
\end{minipage}
\begin{minipage}{0.19\linewidth}
  \includegraphics[width=3.5cm, height=3.0cm]{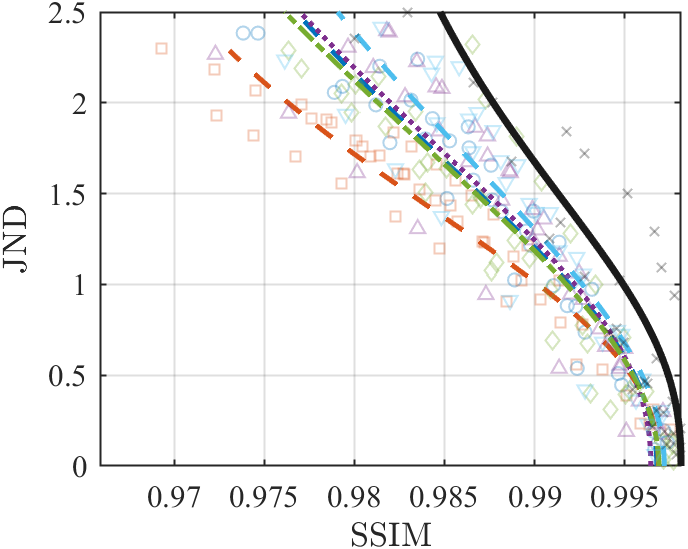}
\end{minipage}
\begin{minipage}{0.19\linewidth}
  \includegraphics[width=3.5cm, height=3.0cm]{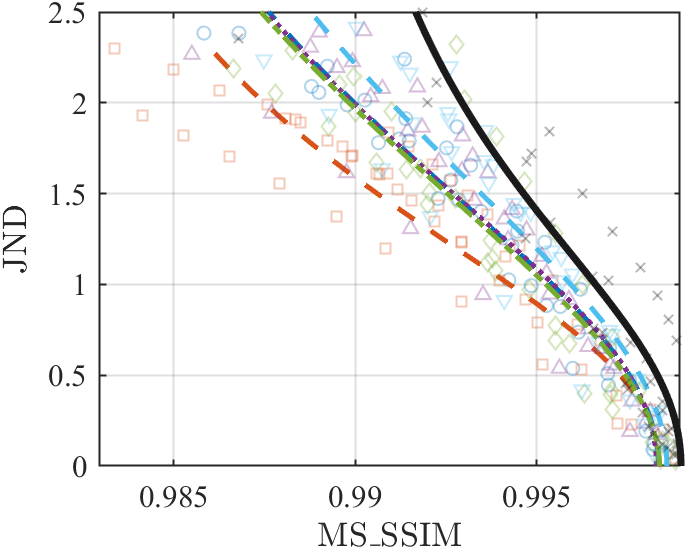}
\end{minipage}
\begin{minipage}{0.19\linewidth}
  \includegraphics[width=3.5cm, height=3.0cm]{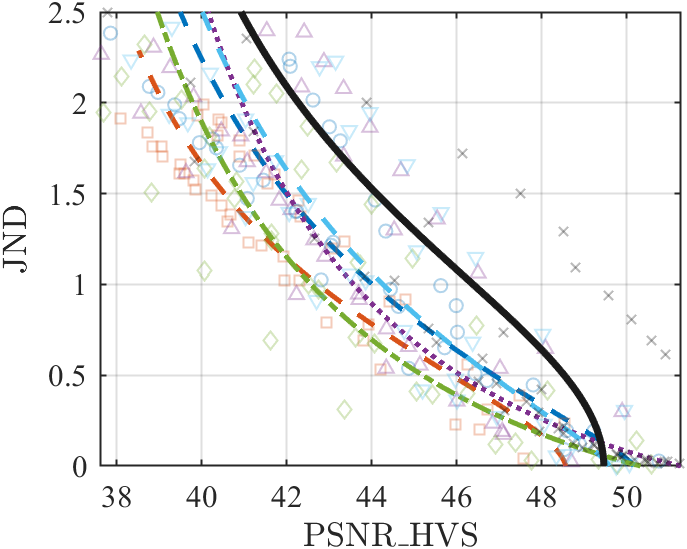}
\end{minipage}
\begin{minipage}{0.19\linewidth}
   \includegraphics[width=3.5cm, height=3.0cm]{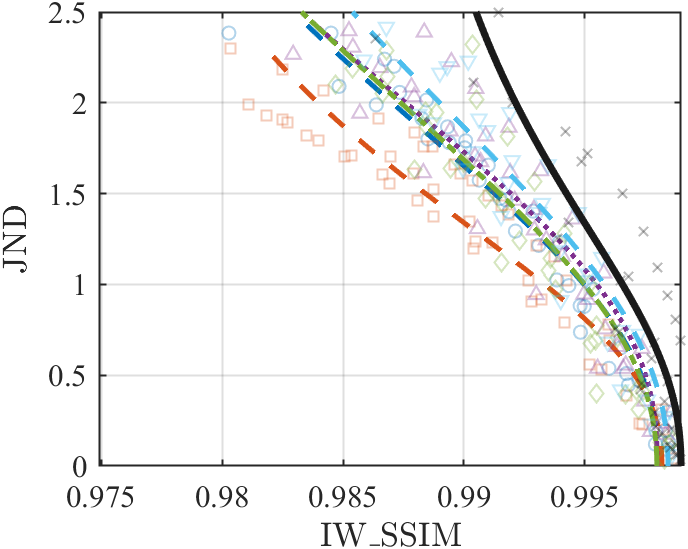}
\end{minipage}
\\
\begin{minipage}{0.19\linewidth}
   \includegraphics[width=3.5cm, height=3.0cm]{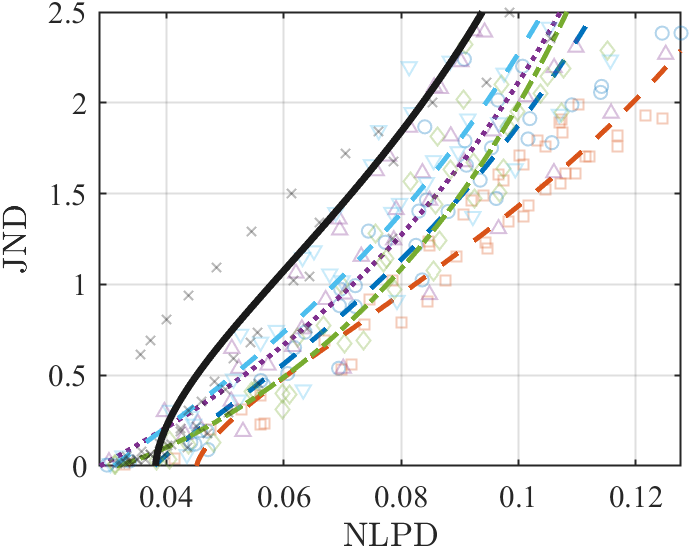}
\end{minipage}
\begin{minipage}{0.19\linewidth}
   \includegraphics[width=3.5cm, height=3.0cm]{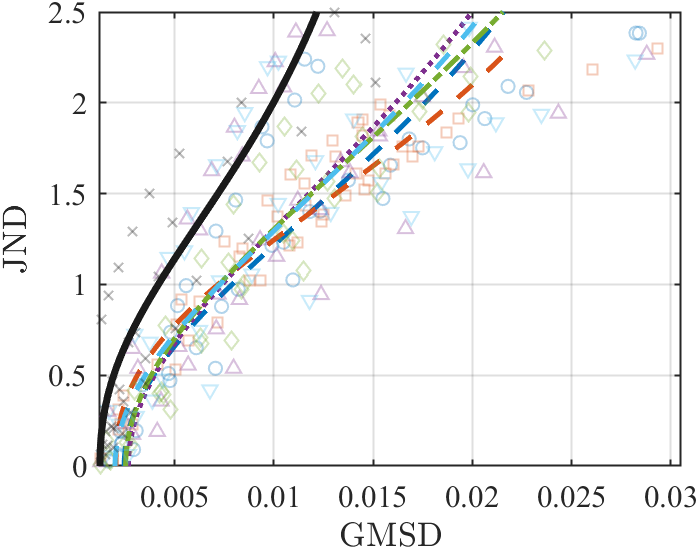}
\end{minipage}
\begin{minipage}{0.19\linewidth}
   \includegraphics[width=3.5cm, height=3.0cm]{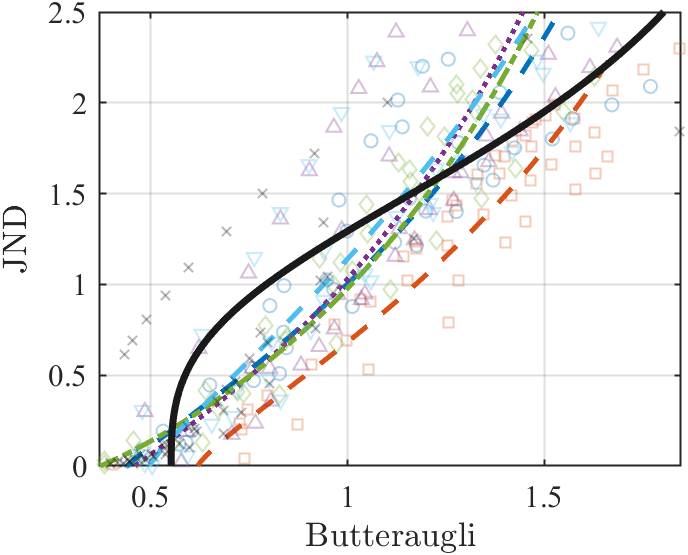}
\end{minipage}
\begin{minipage}{0.19\linewidth}
   \includegraphics[width=3.5cm, height=3.0cm]{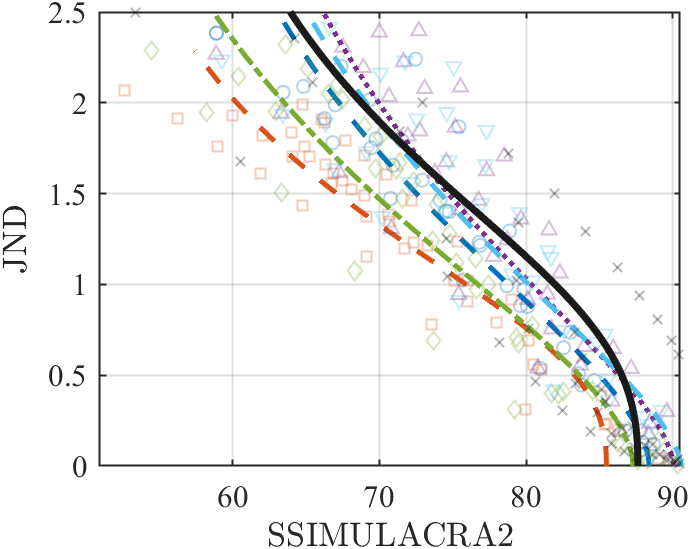}
\end{minipage}
\begin{minipage}{0.19\linewidth}
   \includegraphics[width=3.5cm, height=3.0cm]{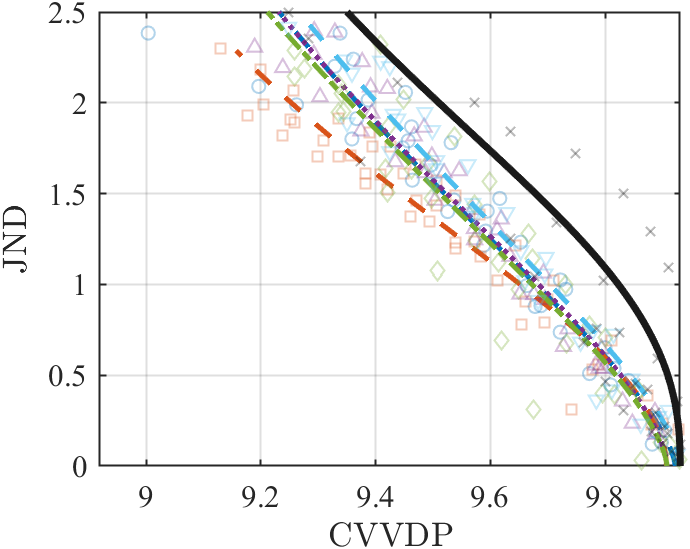}
\end{minipage}
\\
\begin{minipage}{0.19\linewidth}
   \includegraphics[width=3.5cm, height=3.0cm]{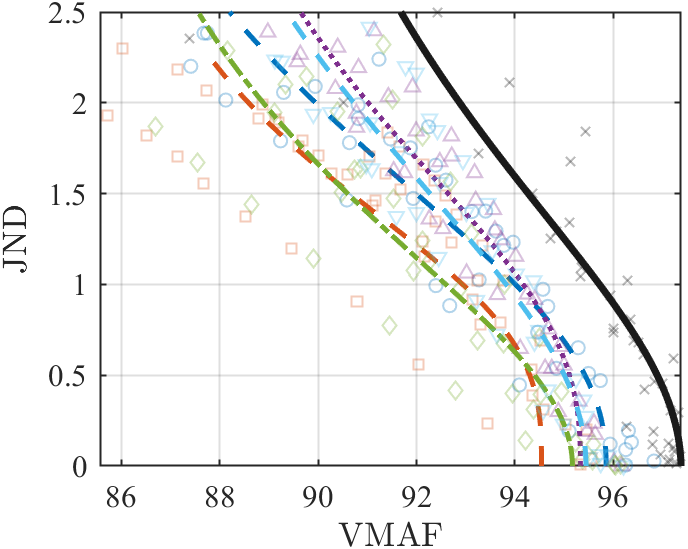}
\end{minipage}
\begin{minipage}{0.19\linewidth}
   \includegraphics[width=3.5cm, height=3.0cm]{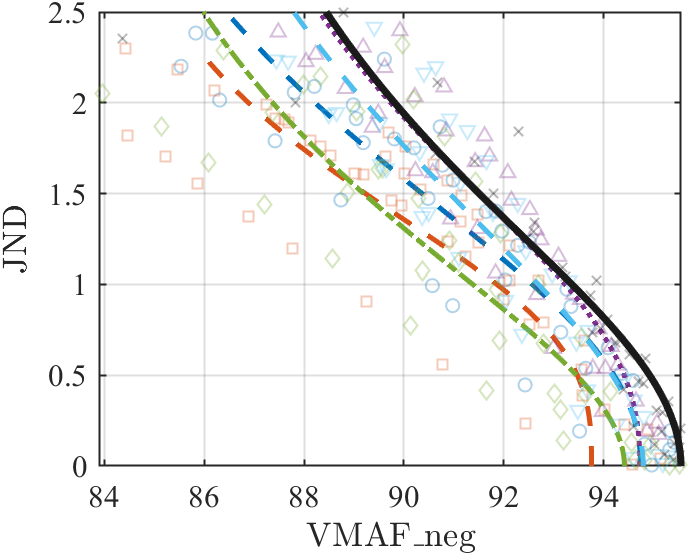}
\end{minipage}
\begin{minipage}{0.19\linewidth}
   \includegraphics[width=3.5cm, height=3.0cm]{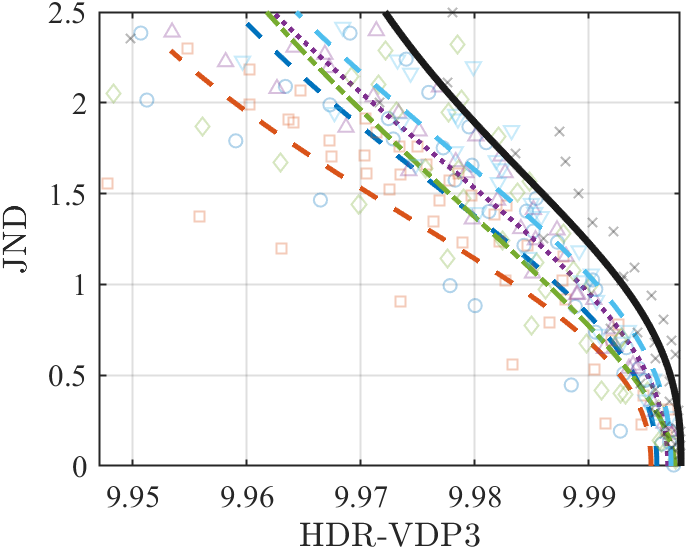}
\end{minipage}
\begin{minipage}{0.19\linewidth}
   \includegraphics[width=3.5cm, height=3.0cm]{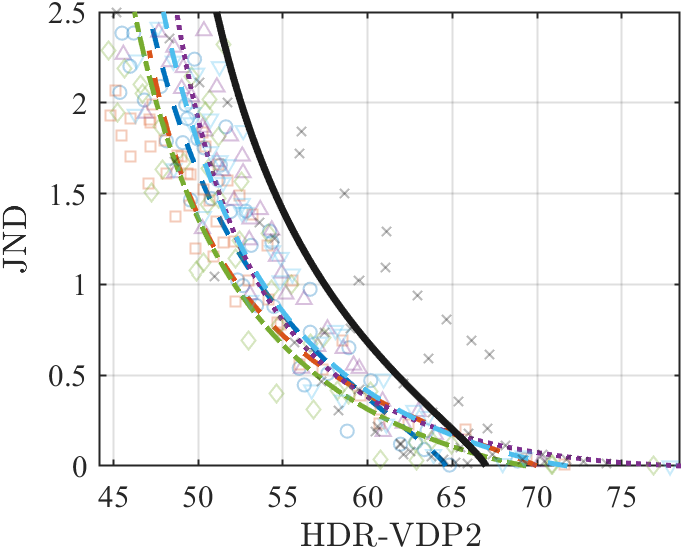}
\end{minipage}
 \begin{minipage}{0.19\linewidth}
\hspace{15pt}
\vspace{5pt}
  \begin{adjustbox}{width=1.6cm, height=2.1cm}    \includegraphics{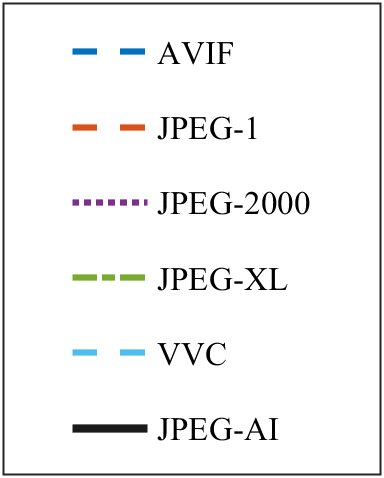}
  \end{adjustbox}
\end{minipage}
\caption{Objective IQA scores are plotted against the corresponding JND values for several metrics. To visualize the trend for each codec, a logistic curve was fitted by minimizing the mean squared error between the metric scores and JND values. The JND values up to 2.5 JNDs are shown for visualization purposes.}
\label{fig:invlogistic_all_eps}
\end{figure*}
\begin{table}[t!]
\centering
\scriptsize
\renewcommand{\arraystretch}{1.15}
\setlength{\tabcolsep}{2pt}
\caption{Absolute correlation values between IQA metric scores and JND values across different aggregation schemes.}
\vspace{-5pt}
\label{tab:metric_correlations}
\begin{tabular}{lcc|cc|cc|cc}
\toprule
& \multicolumn{2}{c|}{\textbf{Overall}} & \multicolumn{2}{c|}{\textbf{Per-source}} & \multicolumn{2}{c|}{\textbf{Per-codec}}& \multicolumn{2}{c}{\textbf{JPEG AI}}\\
\textbf{Metric} & PLCC & SRCC & PLCC & SRCC & PLCC & SRCC & PLCC & SRCC\\
\midrule
PSNR-Y & 0.807 & 0.816 & 0.883 & 0.885 & 0.851 & 0.832 & 0.849 & 0.815  \\
SSIM \cite{ssim} & 0.894 & 0.905 & 0.908 & 0.904 & 0.936 & 0.932 & \underline{0.926} & 0.904  \\
MS-SSIM \cite{ms-ssim} & \underline{0.931} & \underline{0.941} & 0.943 & 0.942 & \underline{0.952} & \underline{0.951} & \underline{0.927} & \underline{0.918}  \\
PSNR-HVS \cite{psnr-hvs} & 0.867 & 0.877 & 0.941 & \underline{0.948} & 0.882 & 0.870 & 0.846 & 0.813  \\
IW-SSIM \cite{iw-ssim} & \underline{0.950} & \underline{0.951} & \underline{0.947} & \underline{0.947} & \underline{\textbf{0.968}} & \underline{\textbf{0.962}} & 0.919 & 0.889  \\
NLPD \cite{nlpd} & 0.900 & 0.913 & 0.927 & 0.927 & 0.936 & 0.935 & 0.923 & \underline{0.905}  \\
GMSD \cite{gmsd} & 0.894 & 0.903 & \underline{0.953} & \underline{0.957} & 0.903 & 0.892 & 0.901 & 0.874  \\
Butteraugli-pnorm \cite{butteraugli} & 0.883 & 0.897 & 0.933 & 0.937 & 0.908 & 0.900 & 0.858 & 0.842  \\
SSIMULACRA1 \cite{ssimulacra1} & 0.898 & 0.908 & 0.907 & 0.902 & \underline{0.955} & \underline{0.957} & \underline{0.923} & \underline{0.904}  \\
SSIMULACRA2 \cite{ssimulacra2} & 0.900 & 0.913 & 0.940 & 0.941 & 0.926 & 0.924 & 0.874 & 0.841  \\
VMAF \cite{vmaf} & 0.882 & 0.891 & 0.903 & 0.900 & 0.912 & 0.925 & 0.855 & 0.899  \\
VMAF-neg \cite{vmaf-neg} & 0.909 & 0.921 & 0.940 & 0.937 & 0.932 & 0.936 & \underline{\textbf{0.958}} & \underline{\textbf{0.963}}  \\
HDR-VDP-2 Q \cite{hdr-vdp-2} & \underline{0.919} & \underline{0.929} & \underline{0.944} & 0.943 & 0.929 & 0.919 & 0.865 & 0.814  \\
HDR-VDP-3 Q \cite{hdr-vdp-3} & \underline{0.919} & \underline{0.933} & \underline{0.948} & \underline{0.950} & \underline{0.937} & \underline{0.939} & \underline{0.931} & \underline{0.953}  \\
CVVDP \cite{cvvdp} & \underline{\textbf{0.960}} & \underline{\textbf{0.961}} & \underline{\textbf{0.962}} & \underline{\textbf{0.962}} & \underline{0.963} & \underline{0.958} & 0.880 & 0.843  \\

\bottomrule
\end{tabular}

\vspace{0.5em}
The top 5 values are \underline{underlined}, and the best is shown in \textbf{bold}.
\vspace{-20pt}
\end{table}

\subsection{Objective metrics evaluation}
We evaluated 15 image quality metrics how well they can predict perceptual quality for high-fidelity image compression. 
Fig.~\ref{fig:invlogistic_all_eps} shows scatter plots of quality metrics versus the perceived distortions. For clarity, we fitted a 4-parameter logistic function to this data per metric and codec by least-squares optimization. 
The plots show that most metrics have a tendency to be more optimistic in predicting the quality of the JPEG~AI compressed images, i.e., they assign better scores (on average) to JPEG~AI images than to images compressed with traditional codecs. One potential explanation for this behavior is that metrics were not designed for the type of artifacts present in learning based compression solutions.

Table~\ref{tab:metric_correlations} compares the performance of the quality predictions by metrics in terms of correlation with the subjective scores from our dataset. The correlations are given in four ways, for the complete dataset of 300 compressed images, per codec, per source, and for the JPEG~AI codec only. We reported both Pearson (PLCC) and Spearman (SRCC) correlation coefficients. The PLCC was computed after mapping each metric’s scores to the subjective JND scores using a  logistic function.

CVVDP has the best performance overall. However, for JPEG~AI images, VMAF-neg provides the best performance. 

Typically in QoE research, the ranking of metrics is decided based on correlation alone. However, this can be improved upon by checking for the statistical significance of the differences of the shown correlations that can be very small. For this purpose, we propose the Meng–Rosenthal–Rubin (MRR) test for dependent correlations~\cite{meng1992comparing}. This test evaluates the null hypothesis that both metrics are equally correlated with the ground truth, that is, \( H_0: r_{XZ} = r_{YZ} \), where \( r_{XZ} \) and \( r_{YZ} \) denote the  correlations of metrics \( X \) and \( Y \) with the subjective scores \( Z \). 
The standardized test statistic is computed as:
\begin{equation}
Z = (z_1 - z_2)\left(\frac{2(1 - r_{XY}) h}{n - 3} \right)^{-0.5},
\end{equation}
where \( z_1 = \tanh^{-1}(r_{XZ}) \), \( z_2 = \tanh^{-1}(r_{YZ}) \), \( r_{XY} \) is the Spearman correlation between metrics \( X \) and \( Y \), and \( n \) is the number of samples. The correction factor \( h \) adjusts for the dependence between predictors:
\begin{equation*}
h = \frac{1 - f \cdot \bar{r}^2}{1 - \bar{r}^2}, \quad
f = \frac{1 - r_{XY}}{2(1 - \bar{r}^2)}, \quad
\bar{r}^2 = \frac{r_{XZ}^2 + r_{YZ}^2}{2}. 
\end{equation*}

A two-tailed test with significance level \( \alpha = 0.05 \) was used to determine whether one metric was significantly more correlated with the subjective scores than another, taking into account all 300 compressed images for the correlations. The absolute Spearman correlation was used, as the sign of the correlation is not relevant in our study. The results of the MRR test are listed in Table~\ref{tab:mrr_results}, where each cell indicates whether the IQA metric in the row is significantly better ({+1}),  worse ({–1}), or not significantly different ({0}) than the metric in the corresponding column.  As expected, perceptually motivated metrics such as CVVDP demonstrate significantly higher correlations with subjective scores compared to traditional metrics like PSNR-Y. This analysis provides statistical confirmation of performance differences.

Note also that even large differences in correlation may not be statistically significant. For example, the overall SRCC for GMSD and VMAF$_\text{neg}$ is 0.903 and 0.921, respectively, but the difference of 0.018 is statistically insignificant here.

\commentout{
\noindent
\scriptsize
\setlength{\tabcolsep}{1pt}
\begin{tabular}{l|cc|cc|cc|cc}
    &  \multicolumn{2}{c|}{Overall} & \multicolumn{2}{c|}{Per-codec (avg)} & \multicolumn{2}{c|}{JPEG~AI} & \multicolumn{2}{c}{Per-source (avg)}\\
metric & PLCC & SRCC & PLCC & SRCC & PLCC & SRCC & PLCC & SRCC \\
\hline
UQI & -0.464 & -0.566 & -0.438 & -0.579 & -0.144 & -0.369 & -0.898 & -0.907\\
NLPD & 0.892 & 0.913 & 0.925 & 0.935 & 0.924 & 0.905 & 0.929 & 0.927\\
GMSD & 0.84 & 0.903 & 0.849 & 0.892 & 0.885 & 0.874 & 0.949 & 0.957\\
DSSIM & 0.877 & 0.925 & 0.899 & 0.942 & 0.903 & 0.915 & 0.921 & 0.939\\
Butteraugli-max & 0.77 & 0.865 & 0.837 & 0.842 & 0.689 & 0.736 & 0.868 & 0.936\\
Butteraugli-pnorm & 0.872 & 0.897 & 0.895 & 0.9 & 0.827 & 0.842 & 0.931 & 0.937\\
SSIMULACRA1 & 0.886 & 0.908 & 0.948 & 0.957 & 0.921 & 0.904 & 0.897 & 0.902\\
SSIMULACRA2 & -0.887 & -0.913 & -0.909 & -0.924 & -0.862 & -0.841 & -0.943 & -0.941\\
CVVDP & -0.96 & -0.961 & -0.963 & -0.958 & -0.88 & -0.843 & -0.962 & -0.962\\
VMAF & -0.874 & -0.891 & -0.901 & -0.925 & -0.864 & -0.899 & -0.901 & -0.9\\
PSNR-Y & -0.785 & -0.816 & -0.827 & -0.832 & -0.829 & -0.815 & -0.881 & -0.885\\
SSIM libvmaf & -0.919 & -0.941 & -0.932 & -0.947 & -0.899 & -0.864 & -0.937 & -0.943\\
MS-SSIM libvmaf & -0.92 & -0.944 & -0.931 & -0.952 & -0.9 & -0.918 & -0.939 & -0.946\\
PSNR-HVS & -0.859 & -0.877 & -0.873 & -0.87 & -0.844 & -0.813 & -0.938 & -0.948\\
VMAF-neg & -0.895 & -0.921 & -0.915 & -0.936 & -0.923 & -0.963 & -0.936 & -0.937\\
PSNR & -0.777 & -0.814 & -0.819 & -0.829 & -0.832 & -0.831 & -0.876 & -0.883\\
SSIM & -0.881 & -0.905 & -0.928 & -0.932 & -0.913 & -0.904 & -0.899 & -0.904\\
MS-SSIM & -0.917 & -0.941 & -0.93 & -0.951 & -0.896 & -0.918 & -0.935 & -0.942\\
IW-SSIM & -0.938 & -0.951 & -0.951 & -0.962 & -0.903 & -0.889 & -0.936 & -0.947\\
IW-PSNR & -0.819 & -0.848 & -0.822 & -0.817 & -0.786 & -0.737 & -0.921 & -0.945\\
HDR-VDP-3 Q & -0.871 & -0.933 & -0.879 & -0.939 & -0.84 & -0.953 & -0.93 & -0.95\\
HDR-VDP-2  & -0.879 & -0.929 & -0.888 & -0.919 & -0.836 & -0.814 & -0.913 & -0.943\\
\end{tabular}

\begin{figure*}
\centering
\includegraphics[width=\linewidth]{figures/metric_bias.pdf}%
\caption{Codec bias in metrics: for each codec, for the JND range covered by all sources, the average metric scores are shown.
The average across all codecs is indicated with a dotted line.
This provides an indication of a codec-dependent bias in the various metrics.}
\label{fig:metric_bias}
\end{figure*}

\begin{figure*}
\includegraphics[width=\linewidth]{figures/metric_correlation_effect.pdf}%
\caption{Metric correlations: comparing correlations for the subset of test images corresponding to the 5 traditional codecs (blue) with correlations for the full set of test images including the JPEG~AI images (black).}
\label{fig:metric_correlation_effect}
\end{figure*}
}

\begin{table}[t]
\centering
\footnotesize
\renewcommand{\arraystretch}{1}
\setlength{\tabcolsep}{1.6pt}
\caption{Meng–Rosenthal–Rubin test result. Each cell shows whether the row metric has a significantly higher (+1), significantly lower (–1), or no significant difference (0) in its correlation with subjective scores compared to the column metric.}
\vspace{-5pt}
\label{tab:mrr_results}
\begin{tabular}{l*{15}{c}}
\toprule
\textbf{Metric} & \rotatebox{90}{PSNR-Y} & \rotatebox{90}{SSIM} & \rotatebox{90}{MS-SSIM} & \rotatebox{90}{PSNR-HVS} & \rotatebox{90}{IW-SSIM} & \rotatebox{90}{NLPD} & \rotatebox{90}{GMSD} & \rotatebox{90}{Butteraugli} & \rotatebox{90}{SSIMU1} & \rotatebox{90}{SSIMU2} & \rotatebox{90}{VMAF} & \rotatebox{90}{VMAF$_\text{neg}$} & \rotatebox{90}{HDR-VDP2} & \rotatebox{90}{HDR-VDP3} & \rotatebox{90}{CVVDP} \\
\midrule
PSNR-Y         & \textcolor{gray}{0} & \textcolor{red}{-1} & \textcolor{red}{-1} & \textcolor{red}{-1} & \textcolor{red}{-1} & \textcolor{red}{-1} & \textcolor{red}{-1} & \textcolor{red}{-1} & \textcolor{red}{-1} & \textcolor{red}{-1} & \textcolor{red}{-1} & \textcolor{red}{-1} & \textcolor{red}{-1} & \textcolor{red}{-1} & \textcolor{red}{-1} \\
SSIM           & {\bf +1} & \textcolor{gray}{0} & \textcolor{gray}{0} & {\bf +1} & \textcolor{red}{-1} & {\bf +1} & {\bf +1} & {\bf +1} & {\bf +1} & {\bf +1} & {\bf +1} & {\bf +1} & {\bf +1} & \textcolor{gray}{0} & \textcolor{red}{-1} \\
MS-SSIM        & {\bf +1} & \textcolor{gray}{0} & \textcolor{gray}{0} & {\bf +1} & \textcolor{red}{-1} & {\bf +1} & {\bf +1} & {\bf +1} & {\bf +1} & {\bf +1} & {\bf +1} & {\bf +1} & {\bf +1} & {\bf +1} & \textcolor{red}{-1} \\
PSNR-HVS       & {\bf +1} & \textcolor{red}{-1} & \textcolor{red}{-1} & \textcolor{gray}{0} & \textcolor{red}{-1} & \textcolor{red}{-1} & \textcolor{red}{-1} & \textcolor{red}{-1} & \textcolor{red}{-1} & \textcolor{red}{-1} & \textcolor{gray}{0} & \textcolor{red}{-1} & \textcolor{red}{-1} & \textcolor{red}{-1} & \textcolor{red}{-1} \\
IW-SSIM        & {\bf +1} & {\bf +1} & {\bf +1} & {\bf +1} & \textcolor{gray}{0} & {\bf +1} & {\bf +1} & {\bf +1} & {\bf +1} & {\bf +1} & {\bf +1} & {\bf +1} & {\bf +1} & {\bf +1} & \textcolor{red}{-1} \\
NLPD           & {\bf +1} & \textcolor{red}{-1} & \textcolor{red}{-1} & {\bf +1} & \textcolor{red}{-1} & \textcolor{gray}{0} & \textcolor{gray}{0} & {\bf +1} & \textcolor{gray}{0} & \textcolor{gray}{0} & {\bf +1} & \textcolor{gray}{0} & \textcolor{red}{-1} & \textcolor{red}{-1} & \textcolor{red}{-1} \\
GMSD           & {\bf +1} & \textcolor{red}{-1} & \textcolor{red}{-1} & {\bf +1} & \textcolor{red}{-1} & \textcolor{gray}{0} & \textcolor{gray}{0} & \textcolor{gray}{0} & \textcolor{gray}{0} & \textcolor{gray}{0} & \textcolor{gray}{0} & \textcolor{gray}{0} & \textcolor{red}{-1} & \textcolor{red}{-1} & \textcolor{red}{-1} \\
Butteraugli-pnorm    & {\bf +1} & \textcolor{red}{-1} & \textcolor{red}{-1} & {\bf +1} & \textcolor{red}{-1} & \textcolor{red}{-1} & \textcolor{gray}{0} & \textcolor{gray}{0} & \textcolor{gray}{0} & \textcolor{red}{-1} & \textcolor{gray}{0} & \textcolor{red}{-1} & \textcolor{red}{-1} & \textcolor{red}{-1} & \textcolor{red}{-1} \\
SSIMULACRA 1         & {\bf +1} & \textcolor{red}{-1} & \textcolor{red}{-1} & {\bf +1} & \textcolor{red}{-1} & \textcolor{gray}{0} & \textcolor{gray}{0} & \textcolor{gray}{0} & \textcolor{gray}{0} & \textcolor{gray}{0} & {\bf +1} & \textcolor{gray}{0} & \textcolor{red}{-1} & \textcolor{red}{-1} & \textcolor{red}{-1} \\
SSIMULACRA 2         & {\bf +1} & \textcolor{red}{-1} & \textcolor{red}{-1} & {\bf +1} & \textcolor{red}{-1} & \textcolor{gray}{0} & \textcolor{gray}{0} & {\bf +1} & \textcolor{gray}{0} & \textcolor{gray}{0} & {\bf +1} & \textcolor{gray}{0} & \textcolor{red}{-1} & \textcolor{red}{-1} & \textcolor{red}{-1} \\
VMAF           & {\bf +1} & \textcolor{red}{-1} & \textcolor{red}{-1} & \textcolor{gray}{0} & \textcolor{red}{-1} & \textcolor{red}{-1} & \textcolor{gray}{0} & \textcolor{gray}{0} & \textcolor{red}{-1} & \textcolor{red}{-1} & \textcolor{gray}{0} & \textcolor{red}{-1} & \textcolor{red}{-1} & \textcolor{red}{-1} & \textcolor{red}{-1} \\
VMAF-neg & {\bf +1} & \textcolor{red}{-1} & \textcolor{red}{-1} & {\bf +1} & \textcolor{red}{-1} & \textcolor{gray}{0} & \textcolor{gray}{0} & {\bf +1} & \textcolor{gray}{0} & \textcolor{gray}{0} & {\bf +1} & \textcolor{gray}{0} & \textcolor{gray}{0} & \textcolor{red}{-1} & \textcolor{red}{-1} \\
HDR-VDP2       & {\bf +1} & \textcolor{red}{-1} & \textcolor{red}{-1} & {\bf +1} & \textcolor{red}{-1} & {\bf +1} & {\bf +1} & {\bf +1} & {\bf +1} & {\bf +1} & {\bf +1} & \textcolor{gray}{0} & \textcolor{gray}{0} & \textcolor{gray}{0} & \textcolor{red}{-1} \\
HDR-VDP3       & {\bf +1} & \textcolor{gray}{0} & \textcolor{red}{-1} & {\bf +1} & \textcolor{red}{-1} & {\bf +1} & {\bf +1} & {\bf +1} & {\bf +1} & {\bf +1} & {\bf +1} & {\bf +1} & \textcolor{gray}{0} & \textcolor{gray}{0} & \textcolor{red}{-1} \\
CVVDP          & {\bf +1} & {\bf +1} & {\bf +1} & {\bf +1} & {\bf +1} & {\bf +1} & {\bf +1} & {\bf +1} & {\bf +1} & {\bf +1} & {\bf +1} & {\bf +1} & {\bf +1} & {\bf +1} & \textcolor{gray}{0} \\
\bottomrule
\end{tabular}
\vspace{-10pt}
\end{table}

\section{Conclusion}
Our subjective quality assessment study confirmed that JPEG AI can achieve perceptually high-fidelity image compression at very low bitrates. Additionally, objective image quality metrics can reliably predict perceptual impairments in images compressed by JPEG AI and other codecs. However, we observed that most metrics tend to overestimate the visual quality of JPEG AI-compressed images when compared to subjective human evaluations. 
While the CVVDP metric was the best overall, other metrics showed superior performance specifically for JPEG AI-compressed images. Therefore, subjective testing remains essential for accurately evaluating codec performance, particularly in the high-fidelity range. If quality metrics are to be ranked by correlation with ground truth, we recommend applying a statistical significance test like the Meng–Rosenthal–Rubin test.

\newpage 
\bibliographystyle{IEEEbib}
\bibliography{ref}

\end{document}

%% file: acronyms.tex
\newacronym{JND}{JND}{Just Noticeable Difference}
\newacronym{2AFC}{2AFC}{2-Alternative Forced Choice}
\newacronym{AFCN}{AFCN}{2-Alternative Forced Choice without Feedback}
\newacronym{AFCF}{AFCF}{2-Alternative Forced Choice with Feedback}
\newacronym{MTurk}{MTurk}{Amazon Mechanical Turk}
\newacronym{HIT}{HIT}{Human Intelligence Task}